# On the structure and topography of free-standing chemically modified graphene

**N R Wilson[1], P A Pandey[1], R Beanland[1], J P Rourke[2], U Lupo[1], G Rowlands[1] and R A Römer[1,3]**

[1] Department of Physics, University of Warwick, Gibbet Hill Road, Coventry, CV4 7AL, UK

[2] Department of Chemistry, University of Warwick, Gibbet Hill Road, Coventry, CV4 7AL, UK

[3] Centre for Scientific Computing, University of Warwick, Gibbet Hill Road, Coventry, CV4 7AL, UK

E-mail: Neil.Wilson@Warwick.ac.uk

**Abstract.** The mechanical, electrical and chemical properties of chemically modified graphene (CMG) are intrinsically linked to its structure. Here we report on our study of the topographic structure of free-standing CMG using atomic force microscopy and electron diffraction. We find that, unlike graphene, suspended sheets of CMG are corrugated and distorted on nanometre length scales. AFM reveals not only long range (100 nm) distortions induced by the support, as previously observed for graphene, but also short-range corrugations with length scales down to the resolution limit of 10 nm. These corrugations are static not dynamic, and are significantly diminished on CMG supported on atomically smooth substrates. Evidence for even shorter range distortions, down to a few nanometres or less, is found by electron diffraction of suspended CMG. Comparison of the experimental data with simulations reveals that the mean atomic displacement from the nominal lattice position is of order 10% of the carbon-carbon bond length. Taken together, these results suggest a complex structure for chemically modified graphene where heterogeneous functionalisation creates local strain and distortion.

## 1.      Introduction

At first glance the structure of graphene is simple; it is a single atom thick two-dimensional honeycomb lattice of carbon. The $sp^2$ bonding between carbon atoms could naively be expected to keep the graphene structure planar. However, experimental [1-7] and theoretical studies [8-10] have shown the presence of in-plane and out-of-plane distortions, and have suggested that these 'ripples' could limit the electrical properties of graphene [8, 11-13]. The surface topography, or rippling, of graphene has been studied on free-standing sheets by transmission electron microscopy (TEM) [1, 6, 7], and on sheets supported on a variety of substrates by scanning probe microscopy (SPM) [3-5]. SPM has been used to probe the remarkable mechanical strength of free-standing graphene, a result of its unique structure and defect-free nature [14]. Suspended graphene sheets are also an interesting example of tethered flexible electronic membranes [10, 11]; understanding their structure requires, and tests, membrane physics more usually associated with soft condensed matter and biological systems.

Fabrication of graphene from graphite requires exfoliation of the individual layers. This was originally achieved by mechanical exfoliation, a process not amenable to scale-up for industrial graphene production. As a result, alternative routes have been developed, including chemically enabled exfoliation [15]. The most widely used technique involves the oxidation of graphite and subsequent exfoliation in water to produce colloidal suspensions of graphene oxide (GO). Current methods for the subsequent reduction of insulating GO yield material that, although conducting, are far from the pristine two-dimensional graphene obtained by mechanical exfoliation [15]. GO is also the starting material for most other chemically modified graphenes (CMGs), which are of interest in their own right [16]. Applications of CMGs include their use in composites that seek to capitalise on the mechanical strength of graphene [17].

Graphene oxide is a heterogeneous material with an ill-defined stoichiometry of approximately $C_4O_2H$ [16]. Many chemical models of GO exist; there is still considerable debate over their validity, but over some aspects a consensus does seem to have been reached. The most prevalent functional groups appear to be 1,2 epoxides and hydroxyls distributed across the sheet, with carboxyl groups at the edges [16]. The degree of oxidation does not appear to be uniform over the sheet, with highly functionalised regions interspersed with more graphitic regions a few nanometres across [18-21]. There is evidence that a graphene-like lattice is retained [22], although this result could reflect a greater sensitivity to the graphitic regions. There has, however, been little work on the topography of chemically modified graphene, despite its importance for understanding how the changes in mechanical and electrical properties of CMG relative to graphene are linked to changes in the physical structure.

Here we report on a combined AFM and TEM study of the topography and structure of chemically modified graphene. We first present the methods, followed in section 3 by an AFM analysis of the topography of supported and suspended graphene oxide sheets. In section 4 we present an electron diffraction study of suspended GO and chemically reduced GO. In section 5 we introduce a theoretical basis from which to analyse these experimental results, and compare the experimental results to numerical simulations. We find significant differences between the topography of graphene oxide and graphene; differences with relevance to the development of chemical models of graphene oxide, and the reduction of graphene oxide to graphene.

## 2.    Methods

A modified Hummers method [23, 24] was used to synthesize graphene oxide from graphite powder. The full details of the procedure are reported elsewhere [22] along with characterisation of the resultant material. In brief, graphite powder was oxidised by $KNO_3$, concentrated $H_2SO_4$ and $KMnO_4$. This mixture was left for five days, before an aqueous solution of $H_2SO_4$ was added followed by $H_2O_2$. The oxidised material was then washed with an aqueous solution of $H_2SO_4$ and $H_2O_2$ resulting in a dark brown slurry. The subsequent dispersion and exfoliation of the graphite oxide to a colloidal suspension of graphene oxide was aided by either mild sonication in an ultrasound bath, or prolonged stirring. The resultant suspension was then vacuum-filtered through a nitrocellulose filter membrane (Millipore, pore size 0.22 μm) to remove any remaining contaminants, and resuspended at the desired concentration. Graphene oxide was chemically reduced by adding 112 μl of ammonia solution (35 wt %, Fisher Scientific) and 17.6 μl of hydrazine solution (62 wt %, Fisher Scientific) to 10 ml of 1 mg ml$^{-1}$ graphene oxide, the mixture was left to stir for 1 hour at 90 °C [25].

Samples for AFM analysis were made by spin-coating (3000 rpm for 45 s) a drop of GO suspension on the substrate of choice. Typically the GO concentration was between 0.1 mg ml$^{-1}$ and 1.0 mg ml$^{-1}$; changing the concentration changed the surface coverage of the GO, with 1.0 mg ml$^{-1}$ giving almost monolayer coverage. Mica and highly oriented pyrolytic graphite (HOPG) substrates were cleaved immediately prior to spin-coating and then used with no further surface treatment. Prior to spin-coating silicon and silicon oxide substrates were cleaned with an oxygen plasma (1 minute at 100 W rf-plasma in 0.6 mBar oxygen atmosphere); this made the surfaces hydrophilic aiding uniform dispersion of the GO.

GO TEM grids were prepared by placing a drop of GO suspension (0.1 - 0.3 mg ml$^{-1}$) on a lacy carbon support grid (Agar Scientific) and allowing it to dry in air. For investigations which required finding the same region of the grid in both the TEM and the AFM, lacy carbon supports on 'finder grids' (Agar Scientific) were used. Chemically reduced graphene was deposited on TEM grids directly from solution as described for GO.

TEM images and selected area electron diffraction (SAED) patterns were taken using a JEOL 2000FX with a Gatan SC-1000 Orius CCD camera. For SAED a 0.6 μm aperture was used, with a small spot size and spread beam to increase the electron coherence length at the sample. Low-temperature measurements were performed using a liquid nitrogen cooled sample holder. Unless otherwise specified, images and diffraction patterns were taken at 200 kV.

AFM was performed in AC mode (also known as tapping mode) using an Asylum Research MFP3D-SA. The AFM tips were silicon with nominal resonance frequency and spring constant of 75 kHz and 3.5 N m$^{-1}$, and manufacturer's stated tip radius of curvature when first used of < 10 nm (MikroMasch, NSC18 NoAl).

Numerical simulations were written in Mathematica and run on the task farm environment in the University of Warwick Centre for Scientific Computing. Details can be found in the supplementary data available online.

## 3.    Topography directly observed by AFM

SPM has been used to characterise the topography of graphene on a variety of substrates, to study the control of ripple texturing in suspended graphene membranes by thermal stress [2], and to probe the mechanical properties of suspended graphene and CMG sheets [14, 26]. Lui *et al.* used high-resolution AFM to study the topography of graphene deposited by mechanical exfoliation onto mica and silicon oxide surfaces [5]. They found that graphene closely follows the topography of the underlying surface, and that graphene on mica was atomically smooth. By contrast, using UHV-STM and AFM Geringer *et al.* observed regular short wavelength ($\lambda \approx 15$ nm) corrugations in mechanically exfoliated graphene which were not induced by the silicon oxide substrate [3]. A similar study by Ishigami *et al.* showed that the lithography required to make graphene devices leaves a layer of photoresist residue which obscures the underlying graphene; after cleaning they found that the graphene shows no intrinsic corrugations but follows the topography of the underlying substrate with slightly decreased roughness [4].

Gómez-Navarro *et al.* reported STM studies of graphene oxide and chemically reduced graphene oxide which showed that the hexagonal, graphene-like, lattice was partially preserved [19]. Paredes et. al. investigated the topography of GO and chemically reduced graphene oxide on HOPG, and found a *"bumpy morphology with feature sizes ranging between 5 and 10 nm"* and a root-mean-square (rms) roughness of 0.108 nm [18]. Recently Sinitskii *et al.* combined molecular dynamics simulations and AFM to study the corrugation of chemically converted graphene (CCG) monolayers; they concluded that the substrate roughness determines the CCG corrugation [27]. Here we present AFM topography measurements of GO on four different substrates, and compare them to the topography of suspended monolayer GO sheets.

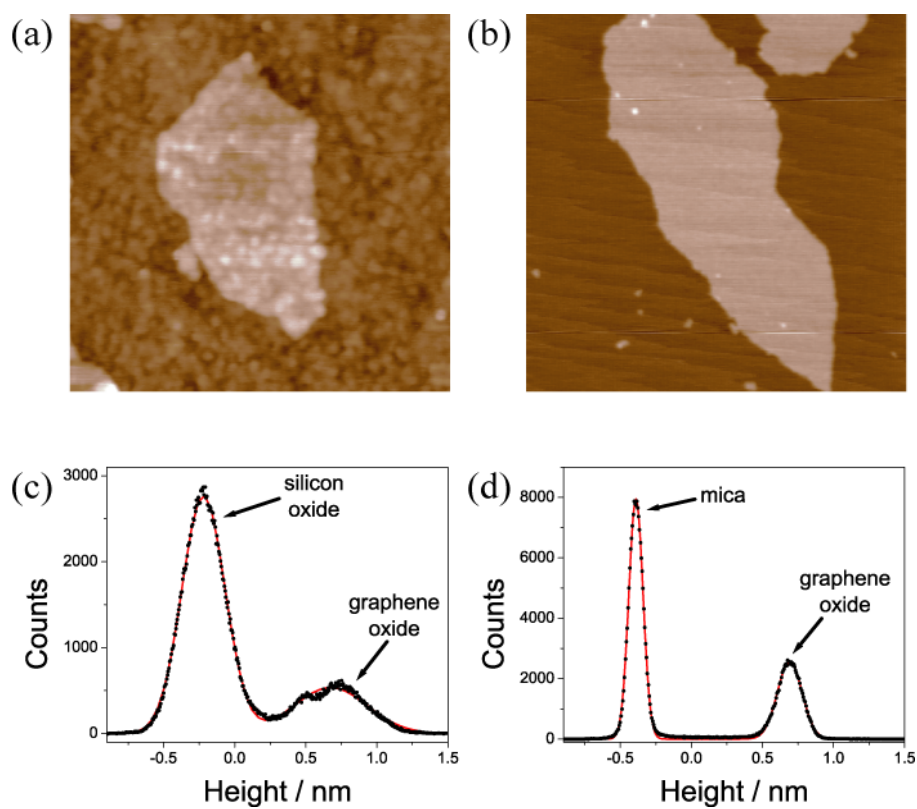

**Figure 1.** 1 μm square AFM topography images of graphene oxide on silicon oxide (a) and mica (b), the full height-scale for both images is 3 nm. (c) and (d) are histograms of the height distributions from (a) and (b) respectively, the black markers are experimental data and the solid red lines are fits to the data as described in the text.

Figure 1 shows 1 μm square AFM topography images of GO on silicon oxide (a) and mica (b), the full height-scale for both images is 3 nm. The images have been post-processed with a second-order flatten using a mask that excludes the GO. Histograms of the height distributions in images (a) and (b) are shown in (c) and (d) respectively. Both histograms show two clear peaks which can be attributed to the substrate and to the GO. The red lines in the histograms are fits to the experimental data using the sum of two Gaussian peaks reflecting the contributions of GO and substrate. The separation between the peaks is the thickness of the GO and the width of the Gaussian fit ($\sigma$) corresponds to the surface roughness. For GO on silicon oxide the observed thickness is 1.08 nm; repeated measurements of many different samples and sheets found a thickness consistently in the range 0.8-1.3 nm, in agreement with previous AFM measurements [18, 19]. The thickness of GO on mica in figure 1 is 0.99 nm, similar to that on silicon oxide. However, the morphology of the GO appears to follow that of the substrate with the rms roughness of GO on silicon oxide significantly greater than GO on mica.

Measurements of the roughness of GO and substrate were taken in this way on samples of GO on mica, on HOPG, on silicon and on silicon oxide. Figure 2 plots rms roughness of the substrate against roughness of the GO. A straight line fit to the data is consistent with a slope of 1, i.e. suggesting that the GO conforms to the underlying substrate. The GO appears to be consistently slightly rougher than the substrate; however, this could be attributable to the post-processing procedure.

A more detailed comparison of the topography of GO and substrate requires images of both to be taken and processed in the same way, and statistical analysis of the results using higher order statistical quantities. Figure 3 shows 500 nm square images of (a) the GO surface on a silicon oxide substrate, and (b) the GO surface on a mica substrate (corresponding images of the substrates are shown in the supporting data available online). Here post-processing involved a third-order flatten of the entire image without masking. Figure 3 (c) and (d) show in red the one-dimensional (1D) height-height correlation function (HHCF) of (a) and (b) respectively, and for the respective substrates in black. The 1D HHCF is a measure of the correlation between the height values of two points as the separation between them increases (see supporting data for further details). The HHCF has been used to analyse e.g. thin film growth morphology [28], self-affine surfaces created by ion-roughening , fluctuating membranes subjected to an external potential [29], and graphene [4, 5]. Here we report only a basic analysis to enable comparison of the topographies. The 1D HHCF is often assumed to be of the form

$$H_x(r_m) = 2\sigma^2 \left(1 - e^{-r_m^2/l_c^2}\right)$$

(1)

where $r_m$ is the (discrete) separation between the points, $\sigma$ is the rms roughness and $l_c$ is the correlation length [28]. Fits of equation 1 to the data are shown as solid lines in figures 3 (c) and (d). Averaged over several such sets of experimental data the roughness and correlation lengths are found to be $\sigma = 0.15 \pm 0.01$ nm and $l_c = 17 \pm 2$ nm for GO on silicon oxide, $\sigma = 0.15 \pm 0.01$ nm and $l_c = 17 \pm 2$ nm for silicon oxide, $\sigma = 0.057 \pm 0.005$ nm and $l_c = 4.7 \pm 0.9$ nm for GO on mica, and $\sigma = 0.033 \pm 0.003$ nm and $l_c = 1.5 \pm 0.4$ nm for mica.

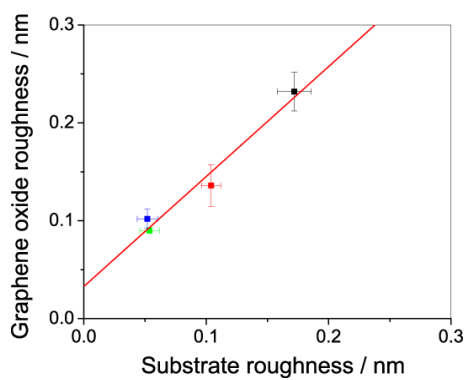

**Figure 2.** Plot of graphene oxide rms roughness versus substrate roughness. The data points correspond to HOPG (blue), mica (green), silicon (red) and silicon oxide (black). Error bars correspond to the spread of data over several measurements of different sheets. The red line is a straight line fit to the data, the best fit slope was found to be $1.1 \pm 0.1$ and intercept $0.03 \pm 0.02$ nm.

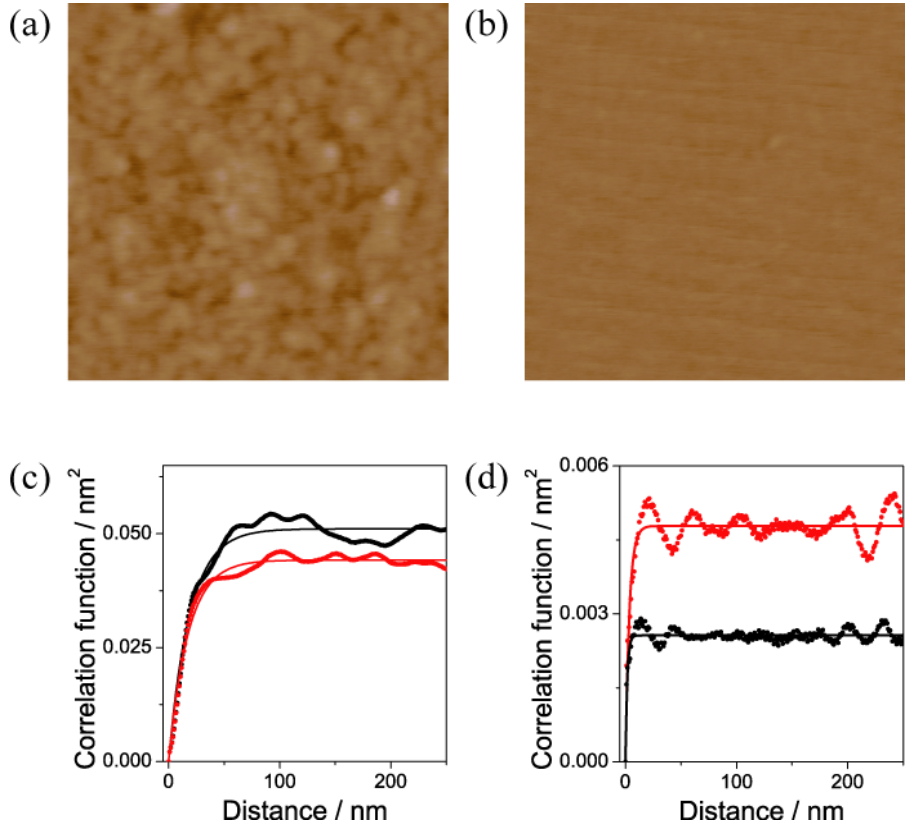

**Figure 3.** 500 nm square AFM height images of (a) graphene oxide on silicon oxide, and (b) graphene oxide on mica; the full height scale for both images is 3 nm. Height-height correlation function analysis of (a) and (b) is shown in (c) and (d) (red points), with data from equivalent images from the respective substrates (black points). The solid lines are fits to the data as described in the text.

The roughness and correlation length of GO on silicon oxide are similar to the silicon oxide itself, further evidence that GO is conforming to the substrate topography as suggested for graphene by Ishigami *et al.* [4]. However, there is a notable difference for GO on mica where the correlation length and roughness increase dramatically relative to the mica itself. The values on mica are attributable to instrumental noise [5], and although we can definitively state that the GO on mica is rougher, the low correlation length of GO on mica suggests that the features are too small to be measured accurately with the AFM tip used here. It is possible that some of the morphology could be due to the presence of adsorbates, however, as these should be equally likely on the mica itself it is probable that the morphology represents features intrinsic to the GO. Assuming that the measured rms roughness measured for mica is due to instrumental noise, and that this adds in quadrature to the substrate roughness, we find the rms roughness of GO on an atomically smooth surface to be $0.024 \pm 0.008$ nm.

The topographic structure of suspended graphene oxide was also examined by AFM. Figure 4 (a) is a bright-field TEM image of GO sheets on a lacy carbon support (dark grey contrast). Overlapping sheets can be seen in the lower and left portions of the image. SAED from the area marked by the dashed red box gave a hexagonal array of diffraction peaks indicative of single sheet graphene oxide (see section 4 and 5). AFM topography images of the whole region and of the area marked by the dashed red box are given in figures 4 (b) and (c) respectively. Figure 4 (c) has been post-processed with a third-order flatten. The 1D HHCF calculated from figure 4 (c) is shown in figure 4 (d) along with a fit of the experimental data using equation 1 (solid red line), giving $\sigma = 0.2$ nm and $l_c = 20$ nm. Repeated images of the same place showed similar features, indicating that the features are static rather than dynamic, are not due to noise and are unlikely to be due to deformation of the substrate by the AFM tip (see supporting data).

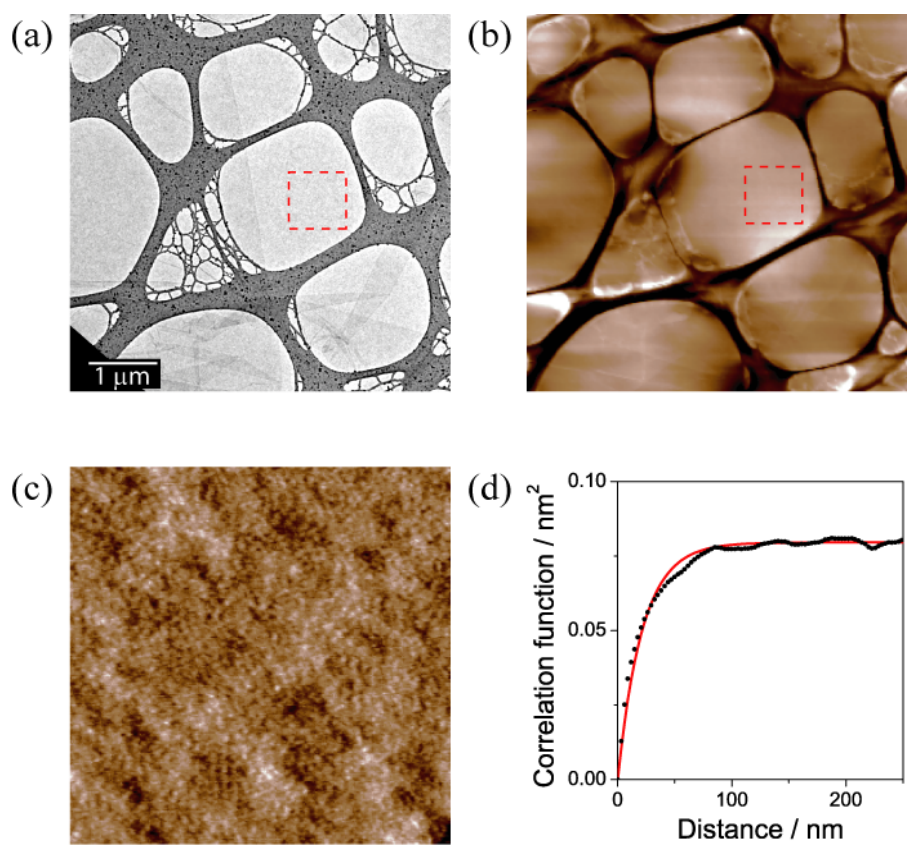

**Figure 4.** (a) bright-field TEM image of graphene oxide on a lacy carbon support. (b) 5 μm square AFM height image of the region as in (a), the full height scale is 35 nm. (c) 750 nm square AFM height image of the area marked by the red dashed square in (a) and (b), the full height scale is 1.5 nm. (d) height-height correlation function analysis of (d) (black points) with fit (red line) as described in the text.

The topography of the suspended GO sheet shows a remarkable amount of structure. In figure 4 (b) distortions of the GO by the lacy carbon support and by uni-directional tension in the sheet can be observed, as also observed for suspended graphene sheets [2]. However, complex structure is also evident in (c). The roughness and correlation length are significantly larger than GO on mica, and slightly larger than GO on silicon oxide, but the morphology is very different. Distortions with length scales as small as 10 nm are evident, comparable to the tip size and hence down to the detection limit of the AFM. The presence of such short range corrugations is perhaps surprising as these would correspond to high degrees of strain in the GO.

Theoretical investigation of tethered flexible membranes, of which the graphene oxide studied here is an example, has predicted functional forms for the HHCF. The short-range behaviour has been shown to be a power-law with exponent dependent on the system; for example a thermally fluctuating membrane has been predicted to have an exponent of 2 [29]. Unfortunately, although the tip radius is larger than the correlation length, and hence measurement of $l_c$ here is reliable, the short-range power-law behaviour is masked by tip-sample convolution effects. This precludes robust determination of the power-law exponent from this data, but demonstrates that if a sharper tip with radius of order 1 nm was used, such as a carbon nanotube tip [30], the power-law behaviour could be studied in detail.

Although the presence and influence of adsorbates on the topography image cannot be ruled out, we expect similar levels of adsorbates in the case of GO on mica; the significantly larger roughness value of the suspended GO sheets suggests that this is not an important factor. It is also unlikely that the features correspond directly to functional groups, but rather to distortions in the graphene lattice induced by them [10]. The topography is consistent with models of GO which include highly functionalised domains interspersed with nanometre scale 'graphitic' regions, and a high degree of strain at the interface between them. Similar corrugations have been inferred in a recent work by Gomez-Navarro *et al.* from HR-TEM images of reduced graphene oxide [31]. However, we believe that this is the first time that their direct observation by AFM on a single suspended GO sheet has been reported.

In summary, AFM measurements show that supported sheets follow the morphology of the substrate to a large degree. When the substrate is atomically smooth the intrinsic roughness of the graphene oxide is observable. Suspended GO is significantly rougher than GO on an atomically smooth substrate; it shows a complex intrinsic structure with corrugations on length scales down to 10 nm, the resolution limit here. Longer wavelength structure is also observed due to interaction between the GO sheet and the lacy carbon support. However, the AFM measurements here cannot reveal the presence or absence of shorter wavelength corrugations due to the comparatively large radius of the AFM tip. The correlation length measured for suspended GO is larger than the AFM tip radius, but not sufficiently so to allow a robust analysis of the power-law behaviour of the short-range height-height correlation function and hence comparison with theory. In section 4 we use electron diffraction to probe the short range structure of CMG.

## 4. Electron diffraction measurements of free-standing CMG sheets

The first experimental measurement of the three-dimensional nature of suspended graphene was reported by Meyer *et al.* in 2007 [6, 7]. They suspended individual sheets of mechanically exfoliated graphene on TEM substrates, and recorded selected area electron diffraction (SAED) patterns from the graphene as the tilt angle between the electron beam and graphene sheet was changed. Analysis of the patterns revealed a linear relationship between the width of the diffraction peaks and the tilt angle, and from this they inferred that the graphene was rippled with a wavelength of between 5 and 25 nm and amplitude of around 1 nm. HR-TEM images have also been analysed to extract the rippling in graphene directly, giving similar values [1]. It should be noted that the electron flux in diffraction is orders of magnitude lower than in HR-TEM imaging making it a less damaging technique. As a result the knock-on damage that is seen in HR-TEM of graphene oxide [31], which complicates interpretation of the images, is not likely to be of importance here.

Graphite oxide and graphene oxide have been studied by TEM imaging and diffraction for many years, including work prior to the initial 'discovery' of graphene; see e.g. [32-35]. Single sheets of graphene oxide give a hexagonal SAED pattern indicating that they contain crystalline material. The diffraction peak positions are indistinguishable from those of graphene [22]. Folds or wrinkles in single sheets superimpose further rotationally misaligned hexagonal patterns, as do multiple sheets. As the number of sheets increases, eventually ring patterns are created due to the absence of crystallographic orientation between the sheets.

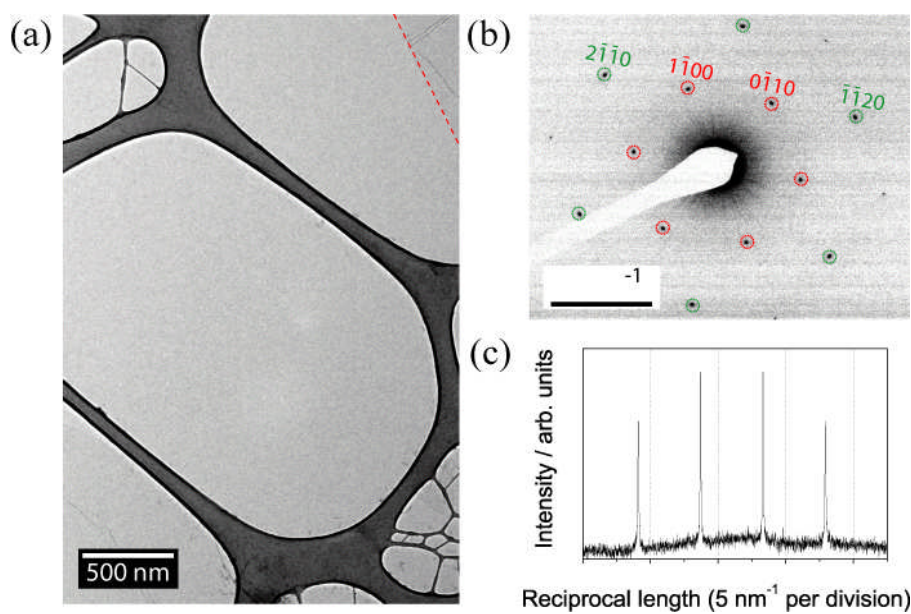

**Figure 5.** (a) Bright-field TEM image of single sheet graphene oxide on a lacy carbon support (dark grey contrast). In the top right corner, marked by a red dashed line, a wrinkled double sheet region can be seen. (b) Selected area electron diffraction taken from the centre of (a), the peaks are indexed and colour coded as described in the text. (c) Plot of the intensity variation along the line defined by the indexed peaks in (b).

Figure 5 (a) is a bright-field TEM image of a single GO sheet suspended on a lacy carbon support grid. The GO is almost completely transparent to the electron beam, as can be seen by comparison to the dark grey lacy carbon support. In the top right of the image, marked by the red dashed line, a second wrinkled layer of GO lies on top of the first.

A SAED taken from the centre of figure 5 (a) is given in figure 5 (b). The hexagonal array of diffraction peaks indicates the graphene-like crystallinity of the GO [22]. The labelled diffraction peaks are indexed using (*hkil*) Miller-Bravais indices. The inner (*hk*) = (*10*) type peaks are marked with dashed red circles, and the (*11*) type peaks with dashed green circles. Some of the (*20*) peaks are also visible (unmarked). Figure 5 (c) is a line section through the indexed peaks in figure 5 (b). It is clear that the inner (*10*) peaks are more intense than the (*11*) peaks, indicative of single layer graphene (see discussion in section 5). Analysis of the diffraction peak intensities reveals significant variations. Typically the ratio of the intensities of the (*10*) type, $I_{10}$, to the intensities of the (*11*) type, $I_{11}$, is found to be between $I_{10} / I_{11} = 1.2$ and $I_{10} / I_{11} = 2$. These values are consistent with previous literature values for single sheet graphene oxide (see section 5). Note that it is important to measure the total scattered intensity associated with the peaks rather than just the peak heights as the peak widths can also vary.

SAED patterns from a different GO sheet are shown in figure 6: (a) is taken at 0° tilt, i.e. with the electron beam perpendicular to the plane of the GO; (b), (c) and (d) are taken at 6° intervals as marked. The approximate tilt axis is indicated by the dashed red line. A (*11*) peak approximately perpendicular to the tilt axis is highlighted by the green arrow in each image. As the tilt angle increases the peak moves outward, indicating a decrease in the projected lattice spacing, and becomes more diffuse.

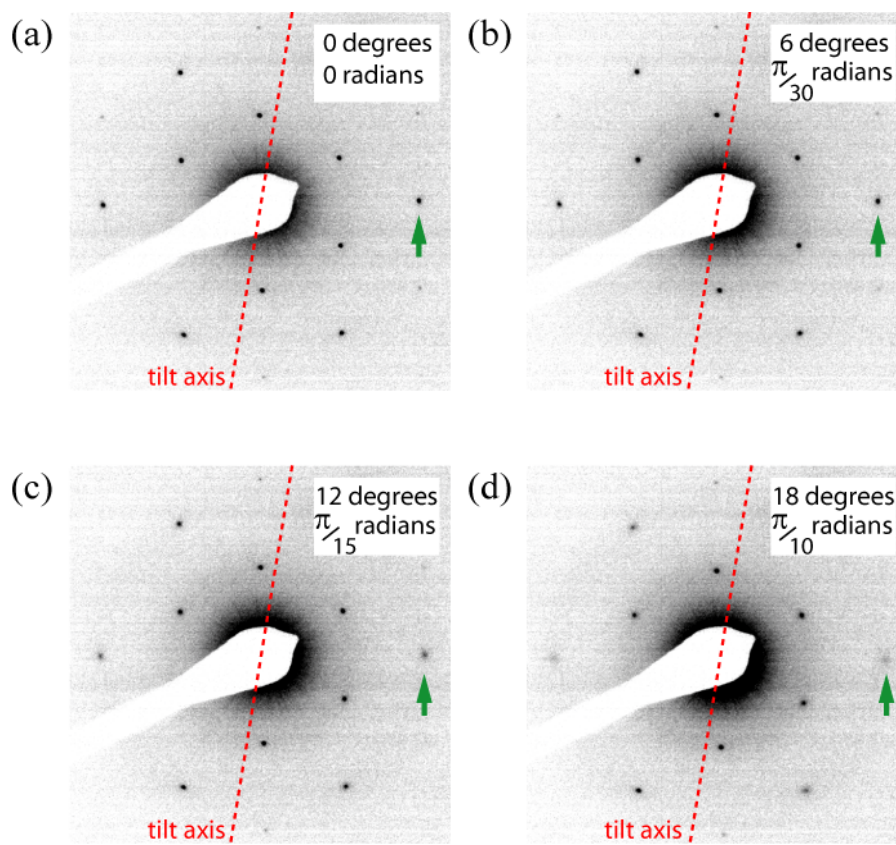

**Figure 6.** Selected area electron diffraction patterns of single sheet graphene oxide taken at (a) 0°, (b) 6°, (c) 12°, (d) 18°. The red-dashed line marks the approximate tilt axis.

The change in peak shape is quantified in figure 7. Line profiles radially outwards through the marked peak are given in figure 7 (a), with experimental data in black, and 1D Gaussian fits in red. The curves have been offset for clarity. The broadening of the peaks is evident, and is quantified in figure 7 (b). Here the width of the diffraction peak has been defined as the standard deviation of the Gaussian fit. The fractional peak width has been scaled by the scattering angle (i.e. the distance of the diffraction peak from the centre of the diffraction pattern). The widths in the tangential direction were similar, indicating an isotropic broadening. The relationship between width and tilt angle is clearly non-linear.

The decreasing intensity of the diffraction peaks, calculated here as the integrated area under the 1D Gaussian fit [36], is plotted as a function of tilt angle in figure 7 (c). The peak intensities have been normalised by the area at 0° tilt.

Different GO sheets show qualitatively the same behaviour but with different rates of broadening - further results are shown in supporting data. Tilt series analysis of double sheet GO is similar to single sheet GO and to chemically converted graphene reduced by exposure to hydrazine [25]. Tilt series analysis on the same sheet of CCG at room temperature and at 133 K also gave similar results, indicating that the observed broadening of the diffraction peaks is not due to thermal effects.

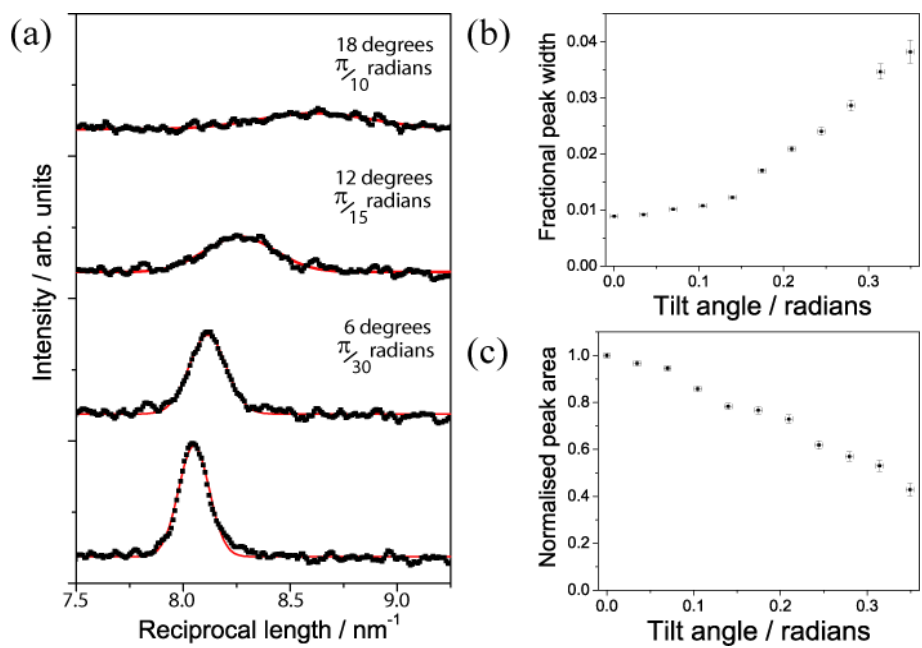

**Figure 7.** (a) Line profiles through the (*11*) diffraction peak labelled with a green arrow in figure 6, the traces are offset for clarity. Black markers are experimental data and solid red lines Gaussian fits as described in the text. (b) Shows a graph of the fractional peak widths, and (c) the normalised peak areas of the Gaussian fits to the (*11*) diffraction peak as a function of tilt angle.

As noted earlier, broadening of the diffraction peaks with tilt angle has been reported for graphene [6, 7], however, the effect here is qualitatively different. The most striking difference is the highly nonlinear dependence of the width on tilt angle, as compared to the linear dependence reported for graphene. The similarity between double and single sheet GO is also in contrast to graphene where double sheets showed reduced peak broadening with tilt angle.

Explaining these results requires further theoretical analysis, and an extension of the numerical simulations reported by Meyer *et al.* [6, 7].

## 5. Analysis

A schematic of a graphene sheet is given in figure 8. The two inequivalent carbon atoms are marked in black and grey, and the lattice vectors $\boldsymbol{a}_1 = \sqrt{3}a_{cc}\left(1/2, \sqrt{3}/2\right)$ and $\boldsymbol{a}_2 = \sqrt{3}a_{cc}\left(-1/2, \sqrt{3}/2\right)$ are in blue (where $a_{cc}$ = 0.142 nm is the carbon-carbon nearest neighbour separation). The ($hk$) = (*10*) and (*11*) lattice planes (lines in this 2D geometry) are marked in green and red respectively. The diffracted intensity, $I$, from a flat 2D graphene sheet ignoring thermal vibrations (i.e. at 0 K) is given by

$$\Phi = f\left(\theta\right)\sum_{n}\exp\left[-i\left(\boldsymbol{k}-\boldsymbol{k}'\right)\boldsymbol{r}_n\right]$$
$$I = \Phi^{*}\Phi \qquad\qquad , \qquad\qquad\qquad (2)$$

where $\boldsymbol{k}$ and $\boldsymbol{k}'$ are the incident and scattered electron beam wavevectors, $\boldsymbol{r}_n$ are the atom positions, and $f(\theta)$ is the scattering factor. The sum is over all atom positions. Diffraction peaks arise at reciprocal lattice vectors $\boldsymbol{k} - \boldsymbol{k}' = \boldsymbol{G}_{hkil}$. The wavelength of the electron beam, $\lambda_e$, is sufficiently small (2.5 pm at 200 kV) that the scattering angle of the ($hk$) peak is given by $\lambda_e/_{d_{hk}} = \sin\theta_{hk} \simeq \theta_{hk}$ [37]. The ratio of the intensities can be calculated from equation (2). For the (*11*) and (*10*) intensities the result is

$$\frac{I_{10}}{I_{11}} = \frac{1}{4}\frac{f(\theta_{10})^2}{f(\theta_{11})^2}.$$  (3)

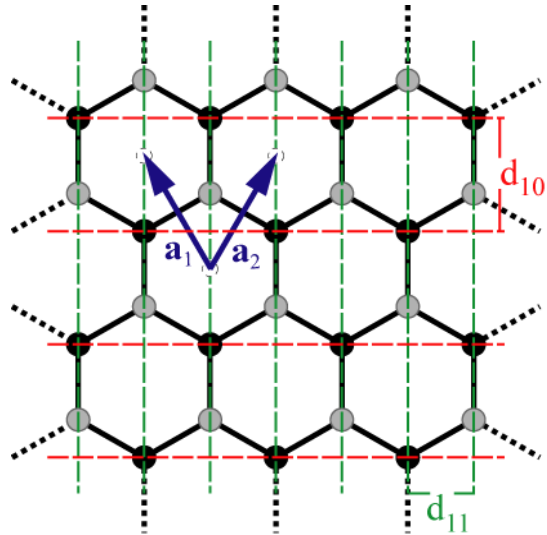

**Figure 8.** Schematic of the structure of graphene. The inequivalent carbon atoms are labelled in grey and black, the lattice vectors in blue, and the lattice planes corresponding to ($hk$) = (*11*) and (*10*) reflections in green and red respectively.

The factor of 1/4 is due to interference between scattering from the inequivalent carbon atoms for the (*10*) lines, as can be seen in figure 8. This effect changes for bi-layer and multi-layer AB (Bernal) stacked graphene. Taking into account the stacking vector, the ratio of (*11*) to (*10*) intensities for N layers of AB stacked graphene / graphite, where *N* is even, is given by

$$\frac{I_{10}}{I_{11}} = \frac{1}{16}\frac{f(\theta_{10})^2}{f(\theta_{11})^2}.$$  (4a)

For odd *N*, writing *N = (2k+1)* the intensity ratio is

$$\frac{I_{10}}{I_{11}} = \frac{k^2+k+1}{4(2k+1)^2}\frac{f(\theta_{10})^2}{f(\theta_{11})^2}.$$  (4b)

For 200 kV electrons, using the scattering factor for carbon from Doyle and Turner [38], this gives for 1,2,3,4,5... large *N*: $I_{11}$ / $I_{10}$ = 1.12, 0.28, 0.37, 0.28, 0.34 ... 0.28 respectively. It is thus possible to distinguish between single layer and multilayer graphene by this ratio alone. The experimentally observed ratio of between 1.2 and 2 is consistent with single layer rather than bilayer or greater, but is much larger than this predicted value. The difference is significant and indicates that atomic displacements of from the nominal 2D lattice sites cannot be neglected.

At finite temperature, thermal oscillations displace the atoms from their equilibrium positions and hence modify the intensities of the diffraction peaks. The Debye-Waller factor accounts for this effect, modifying the scattering factor by a term exponentially dependent on the mean square atomic displacements from their equilibrium positions [39]. We note that the interaction time of an electron in the electron beam with the graphene sheet is far shorter than the thermal oscillation period (frozen phonon approximation); the Debye-Waller analysis is equally valid for thermal oscillations and disordered static displacements. The Debye-Waller factor modifies the intensity ratios; for example the ratio between the (*11*) and (*10*) intensities becomes

$$\frac{I_{10}}{I_{11}} = \frac{I_{10}{}^0}{I_{11}{}^0}\exp[-\frac{16\pi^2\Delta r^2}{(d_{10}{}^2-d_{11}{}^2)}]$$  (5)

where $I_{hk}{}^0$ are the calculated zero temperature intensities, and $\Delta r$ is the mean squared displacement of the atoms from their equilibrium lattice positions. As noted by Horiuchi *et al.* [35] the experimental values for $I_{10}$ / $I_{11}$ can be compared with the predicted (0 K) values to extract a mean square displacement. Using $I_{10}$ / $I_{11}$ = 1.3 ± 0.1 (as measured from the diffraction pattern in figure 5) a value of $\Delta r = 0.010 \pm 0.002$ nm can be calculated.

This value of $\Delta r$ can be used to predict the ratio between the (*11*) peaks and the higher order peaks as shown in Table 1, which compares the theoretical graphene 0 K intensity ratio ($I^0$) with the experimentally observed ($I^{exp}$) values and those calculated using $\Delta r = 0.010$ nm ($I^{calc}$).

**Table 1.** Comparison of the intensity ratios for the (*hk*) peaks: predicted zero-temperature ($I_{hk}{}^0$), experimentally observed ($I_{hk}{}^{exp}$), and predicted using $\Delta r = 0.010$ nm ($I_{hk}{}^{calc}$).

| (*hk*) | $d_{hk}$ / nm | $\theta_{hk}$ / radians | $I_{hk}{}^0/I_{11}{}^0$ | $I_{hk}{}^{exp}/I_{11}{}^{exp}$ | $I_{hk}{}^{calc}/I_{11}{}^{calc}$ |
|---|---|---|---|---|---|
| (*10*) | 0.213 | 0.0118 | 1.12 | 1.3 ± 0.1 | 1.3 ± 0.1 |
| (*11*) | 0.123 | 0.0204 | 1.00 | 1.00 | 1.00 |
| (*20*) | 0.107 | 0.0236 | 0.15 | 0.11 ± 0.03 | 0.14 ± 0.01 |
| (*21*) | 0.081 | 0.0312 | 0.06 | 0.08 ± 0.03 | 0.04 ± 0.01 |
| (*30*) | 0.071 | 0.0353 | 0.14 | 0.05 ± 0.05 | 0.09 ± 0.07 |

The correlation between $I^{exp}$ and $I^{calc}$ suggests that the observed intensities are consistent with displacements of the atoms away from their mean position. However, the uncertainty in the intensities of the higher order peaks is too large to be definitive, in particular over whether the displacements are random or not. The calculated $\Delta r$ for the observed range of intensity ratios is between 5% and 13% of the C-C bond length. This is much larger than would be expected due to thermal oscillations alone. The derived value of $\Delta r$ must thus be interpreted as distortions of the lattice at the atomic scale. These are likely to be due to functionalisation of the graphene and to defects that have arisen as a result of the functionalisation. The variation in the observed intensity ratios can thus be interpreted as variations in the degree of functionalisation, consistent with other reports [16].

The absolute values of the displacements calculated in this way should be treated with caution: graphene oxide does not only contain a graphene-like carbon lattice. The oxygen functional groups also scatter the electron beam; although they have no order (which would be apparent as a superlattice in the diffraction pattern) their regular positioning relative to the carbon atoms leads to changes in the diffracted intensities. However, $\Delta r \sim 0.1\ a_{cc}$ can be taken as an indicative value of the degree of distortion.

### 5.1. Tilted lattice

The analysis so far has been for the electron beam perpendicular to the graphene plane. The tilted geometry is shown in the schematic in figure 9 (a). At a tilt angle of $\alpha$ the projected lattice spacing is decreased resulting in an increase of the scattering angle to

$$\theta_{hk}(\alpha) = \frac{\lambda_e}{d_{hk}\cos\alpha}$$

(6)

If the lattice is perfectly flat the diffraction peak moves outwards with increasing $\alpha$ according to this relationship and the diffracted intensity decreases due to modulation by the scattering factor, but the peak width stays the same. The observed broadening of the diffraction peaks is thus further clear evidence of the displacement of atoms away from 2D lattice sites.

To incorporate the effect of these corrugations the relevant length scales must first be introduced. The wavelength of the electron beam $\lambda_e << a_{cc}$, whilst the coherence length of the electron beam is typically $l_c \sim 1$-10 nm i.e. $\sim 10$ - $100\ a_{cc}$. The size of the aperture used here for SAED is 0.6 μm, i.e. > 4000 $a_{cc}$. The GO sheet is suspended across a hole typically ~ 1 μm in width.

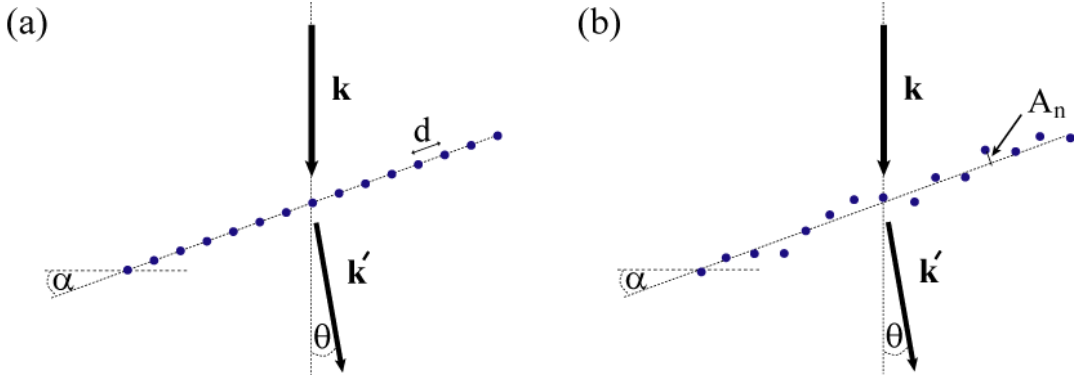

**Figure 9.** Schematic of the diffraction geometry for a tilted 1D (a) flat and (b) rippled lattice.

Considering the distortions as ripples with wavelength $\lambda_r$, in the context of the SAED results long-wavelength ripples are defined as those for which $\lambda_r \gg l_c$ whilst short wavelength ripples are those for which $\lambda_r < l_c$. For length scales less than the coherence length scattering must be added coherently (i.e. adding $\Phi$), for length scales larger than $l_c$ scattering is added incoherently (i.e. adding $I$). As the aperture is much larger than $l_c$ the experimental situation can be viewed as summing the intensities of many different sections of lattice of size $l_c$.

For long wavelength ripples the change in gradient within a coherence length is small; the lattice sections can be considered to be locally flat but tilted. The sum over lattice sections thus includes a range of effective tilt angles, $\delta\alpha$, corresponding to the local gradients. According to equation (6) these will have different $\theta_{hk}$ resulting in a total diffraction peak of width $\delta\theta_{hk}$

$$\frac{\delta\theta_{hk}}{\theta_{hk}} = \frac{1}{\theta_{hk}} \frac{\partial\theta_{hk}}{\partial\alpha} \delta\alpha = \tan\alpha \; \delta\alpha$$

(7)

At small tilt angles this predicts a linear broadening with slope $\delta\alpha$, as reported for graphene [6, 7]. However, this does not fit the experimental data reported here for graphene oxide.

### 5.2. Short wavelength distortions

For short wavelength ripples coherent scattering must be included and the distorted and tilted lattice positions inserted into equation (2). We consider distortions out of the plane of the graphene sheet. For simplicity and to minimise computational expense a 1D lattice is used where the 'atoms' on the 1D lattice represent the lattice planes in 2D. Incorporating the effect of tilt, as shown schematically in figure 9 (b), and expanding to first order in $\theta$, equation (2) becomes

$$\Phi = f(\theta) \sum_n \exp\left[-i\left(nd\cos\alpha - A_n\sin\alpha\right)k\theta\right]$$

(8)

where $A_n$ is the out-of-plane displacement of the $n$th atom and $k$ is the magnitude of the electron beam wavevector. It can be shown (see supporting data) that if $A_n$ is a sinusoidal distortion of wavelength $\lambda_r < l_c$ and amplitude $A$ two satellite peaks appear next to the main diffraction peak with amplitude proportional to $A \sin(\alpha)$ and spacing inversely proportional to $\lambda_r$.

No satellite peaks are observed in the experimental results. However, considering the real lattice distortion as a Fourier series, a range of $\lambda_r$ values will lead to overlapping satellite peaks and hence the appearance of a single broad peak that increases in apparent width with tilt angle as the amplitudes of the overlapping satellite peaks increase.

We have not found an analytical solution to equation (8) with a Fourier expansion of $A_n$ and so instead turn to numerical simulations. Further details are included in the supporting data. The distortion to the lattice is defined by

$$A_n = \sum_i A_i \sin\left( \frac{2\pi nd}{id + \lambda_c} + \phi_i \right) \tag{9}$$

where $A_i$ and $\phi_i$ are (positive) random numbers defining the amplitude and phase of the given wavelength in the Fourier series. The cut-off wavelength, $\lambda_c$ is a parameter which describes the minimum wavelength of the distortion. The average amplitude and hence average distortion of the sheet is constrained by a second parameter, $A_r$, such that the mean $\langle A_i \rangle = A_r$. The coherence length, $l_c$, of the electron beam was introduced by modulating equation (8) by an exponential factor which created a diffraction window of width $l_c / \cos \alpha$ (incorporating the effect of tilt angle). Each iteration creates a 1D lattice distorted by a set of random parameters as described by equation (9), and calculates the resultant diffracted intensity as a function of scattering angle $\theta$, and tilt angle $\alpha$. To model the total scattering intensity, i.e. including incoherent summation of intensity across the aperture, each simulation was averaged over at least 100 iterations. For the simulations shown here $d = d_{11}$.

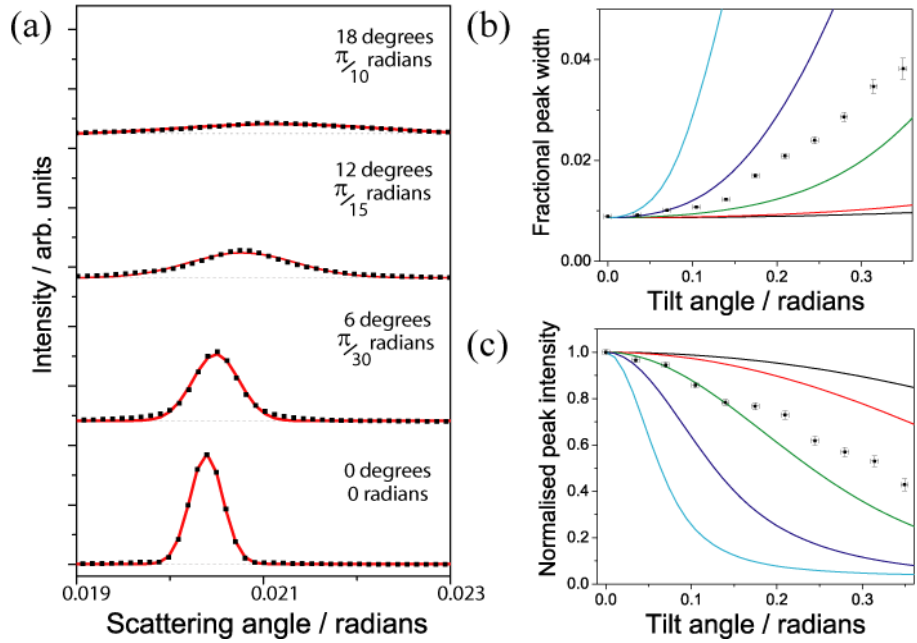

**Figure 10.** (a) Line profiles through the simulated (*11*) diffraction peak with $A_r = 0.1$ *d*, $l_c$ = 5.0 nm and $\lambda_c = 1$ *d*. The traces are offset for clarity. Black markers are simulation data and solid red lines the associated Gaussian fits. (b) Shows a graph of the fractional peak widths, and (c) the normalised peak areas of the Gaussian fits to the simulated diffraction data as a function of tilt angle. Plotted are simulated data for $A_r = 0.01$ (black), 0.02 (red), 0.05 (green), 0.1 (dark blue) and 0.2 (light blue) *d* with $l_c$ = 5.0 nm and $\lambda_c = 1$ *d*; the experimental data from figure 7 is plotted as black markers.

Figure 10 (a) shows line profiles of intensity versus scattering angle as a function of tilt angle for $A_r = 0.1$ *d*, $l_c$ = 5.0 nm and $\lambda_c = 1$ *d*. The results are diffraction peaks with no satellites, but instead decreasing amplitude and increasing width with tilt angle. The simulated data is represented as black markers and the red lines are Gaussian fits to the data. The curves have been offset for clarity. As figure 10 (a) shows, the simulated diffraction peaks are close enough to Gaussian in shape that the peak heights and widths extracted from the Gaussian fits are a fair representation of the diffraction peak characteristics, as was the case for the experimental results. We note that in principle accurate determination of the experimental diffraction peak shape could be used to determine the shape of the graphene / graphene oxide surface. This would require a significant increase in the signal to noise ratio in the experimental data, and an extension of the analysis presented here.

As in the experimental data, figure 7 (a), the peak becomes broader and decreases in intensity as the tilt angle decreases. This is quantified in figure 10 (b) and (c) where the width and peak height of the Gaussian fits to the simulated data is plotted as a function of tilt angle. Plotted as solid lines are the results for $A_r = 0.2$, 0.1, 0.05, 0.02, 0.01 $a_{cc}$, $l_c$ = 5.0 nm and $\lambda_c = d$, with the experimental data from figure 7 (b) plotted as black markers. The qualitative similarities between the experimental data and the simulations are clear; quantitatively the data lies between $A_r = 0.1$ and 0.05 $a_{cc}$. These correspond to mean square displacements of atoms from their nominal lattice positions of between 0.12 and 0.06 $a_{cc}$ [40].

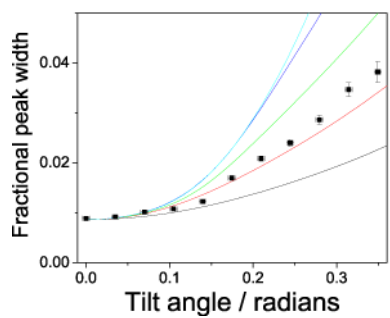

**Figure 11.** Plot of the fractional peak widths of the Gaussian fits to the simulated diffraction data as a function of tilt angle. Plotted are simulated data for $A_r = 0.1$ $d$ with $l_c$ = 5.0 nm and $\lambda_c$ = 1 (light blue), 10 (dark blue), 20 (green), 50 (red), and 100 (black) $d$; the experimental data from figure 7 is plotted as black markers.

The coherence length sets the minimum width of the peaks, i.e. the width at zero tilt. Comparison with the simulated data shows that the experimental coherence length in this instance is greater than 5 nm. Other factors such as spherical aberration can influence the width of the diffraction peaks so it is difficult to accurately determine the experimental coherence length in this way, but a minimum value can be defined. Extending this argument, the zero tilt experimental peak width implies that the graphene oxide has a graphene-like crystalline lattice over length-scales of at least 5 nm.

We can also investigate the effect of the length scale of the distortions by varying the cut-off wavelength $\lambda_c$. Figure 11 plots the widths of the Gaussian fits to the simulated data for $A_r$ = 0.1 $a_{cc}$, $l_c$ = 5.0 nm and $\lambda_c$ = 1, 10, 20, 50, 100 $d$. A cut-off of 100 $d$ is of order the coherence length and hence, as expected, the widths increase almost linearly with tilt angle. As the cut-off wavelength decreases, the peak widths increase more rapidly at larger tilt angle. Interestingly the experimental data seems most consistent with a cut-off wavelength of between 20 $d$ and $50\ d$, which corresponds to a length of between 2.5 nm and 5.0 nm. This is consistent with recent reports which suggest that graphene oxide consists of nanoscale domains of less functionalised 'graphitic regions' interspersed amongst highly functionalised domains.

The analysis developed here demonstrates that the SAED tilt series analysis of GO is consistent with distortions of the graphene-like lattice on nanometre length scales or less. The size of the distortions corresponds to mean atomic displacements of around 10 % of the inter-atomic spacing, i.e. corresponding to large strain in the lattice. This contrasts to the reported results for graphene and indicates fundamental differences in the structure and topography of chemically modified graphene compared to graphene itself.

## 6.    Conclusions

We have used AFM and electron diffraction to investigate the structure of graphene oxide. This reveals a complex topography with distortions and structural features at different length-scales.

AFM shows that when in intimate contact with a rough surface, GO follows the topography of the surface, with no evidence for length scales intrinsic to the GO. However, on atomically smooth surfaces such as mica, the intrinsic roughness of GO (approximately 20 pm) is evident. AFM of suspended GO shows long wavelength (>100 nm) distortions which can be attributed to tension induced by the support material, but also short range corrugations with length scales down to 10 nm. Statistical analysis of the topography of suspended GO shows a roughness of 0.2 nm and a coherence length of 20 nm, significantly greater than the values found for GO on mica and indicative of the complex structure of suspended GO. Analysis at shorter length scales is limited by the ~ 10 nm tip radius.

Analysis of electron diffraction by suspended single-layer CMG sheets reveals clear evidence for corrugations in GO distinct from those previously reported for graphene [6, 7]. Comparisons with numerical simulations and analytical theory indicate that the distortions occur on length scales of a few nanometres or less and correspond to atomic displacements of order 10 % of the carbon-carbon bond length. Such high strain, short range distortions have a dominant effect on the electron diffraction compared to the longer range, lower strain distortions previously reported for graphene. The short-range corrugations are also present in chemically reduced graphene and in bilayer graphene oxide and are too large to be thermal oscillations. Instead, they can tentatively be attributed to strain in the lattice created by functional groups and are consistent with recent reports suggesting a complex heterogeneous structure for graphene oxide [18-21].

Both AFM and electron diffraction clearly indicate that the topography of graphene oxide is different to that reported for graphene. The local strain induced by functionalisation, and in particular by heterogeneous functionalisation, manifests itself as short range corrugations in the graphene-like carbon lattice of graphene oxide. This local strain has important ramifications for the mechanical properties of chemically modified graphene, and for its reactivity. Current simulations of graphene oxide are limited to relatively small numbers of atoms, typically 8 to 128 [41-43]. This work indicates that in order to accurately model the complexity and heterogeneity of graphene oxide, length-scales up to tens of nanometres must be considered.

Reduced graphene oxide appears to exhibit similar corrugations to graphene oxide, consistent with observations that the conductivity of the chemically reduced material is orders of magnitude lower than exfoliated graphene. Combining thermal annealing and chemical reduction has been reported to significantly lower the conductivity of the resultant material [15]. It will be interesting to investigate the extent to which the distortions in the CMG lattice change after thermal annealing and hence the extent to which an intrinsic graphene lattice is recovered. Detection of these changes requires techniques that are sensitive to sub-atomic distortions on nanometre length scales; our work here demonstrates that electron diffraction and AFM can fulfil these requirements.

**Acknowledgements**

We thank Ana Sanchez and Jeremy Sloan for helpful discussions and Steve York for technical support. PAP thanks the Midlands Physics Alliance Graduate School for a scholarship. RB and NRW acknowledge support from the Warwick Centre for Analytical Science (EP/F034210/1). The Gatan Orius digital TEM camera used in this research was funded by Birmingham Science City: Creating and Characterising Next Generation Advanced Materials, with support from Advantage West Midlands and part funded by the European Regional Development Fund.

**Supporting data available**

A video of the diffraction pattern change with tilt angle is available, as is further information on the height-height correlation function and an analytical approximation to diffraction from a 1D lattice with a sinusoidal ripple. Further experimental data for electron diffraction from single and double sheet graphene oxide and from reduced graphene oxide at room temperature and 133 K is also presented, and details about the numerical simulations.